\documentclass{article}
\RequirePackage{graphicx}
\usepackage{graphicx,epsfig}
\usepackage{subcaption}
\usepackage[margin=1in]{geometry}
\usepackage{amsmath}
\usepackage {amssymb}
\usepackage[T1]{fontenc}
\usepackage[utf8]{inputenc}
\usepackage{lineno}
\usepackage[pdftex]{color}
\usepackage{lmodern}
\usepackage{authblk}
\usepackage{blindtext}

\begin{document}

\title{Thermodynamics and Bouncing Cosmology in Rastall-like Gravity
}


\date{}

\author[1]{José A. C. Nogales\thanks{Corresponding Author: jnogales@ufla.br}}
\author[1]{K. Luz-Burgoa\thanks{karenluz@ufla.br}}
\author[2]{Laysa G. Martins\thanks{laysamartinsymail@yahoo.com.br}}

\affil[1]{Departamento de F\'isica  \\ Universidade Federal de Lavras, Lavras UFLA-MG \\ Caixa postal 3037, CEP 37000-900, Brasil.}
\affil[2]{Departamento de Formação Geral, Centro Federal de Educação Tecnológica de Minas Gerais, Varginha, Minas Gerais, CEP 37022-560, Brazil.}

\date{ }

\maketitle

\begin{abstract}
In this study, we explore the thermodynamic aspects of a modified version of Rastall’s gravity theory and its implications for cosmological scenarios. We analyze the role of non-conserved energy-momentum tensor equations and investigate their influence on particle production within an irreversible thermodynamic framework. By introducing a novel Lagrangian, we derive modified field equations and establish their relationship with matter production, both with and without entropy generation. Our analysis focuses on ideal fluid models and extends to spatially flat LFRW cosmologies, providing key equations that govern energy density, pressure, and curvature dynamics. Furthermore, we propose a bouncing cosmological model, in which the universe undergoes cycles of contraction and expansion, avoiding the singularity associated with the Big Bang. Our results indicate that this bouncing scenario is feasible within the Rastall-like gravity framework, supported by particle production processes and stability conditions. The violation of energy conditions near the bounce point further confirms the consistency of this alternative cosmological model. The present work is focused on the theoretical foundations and internal consistency of the model; possible observational implications will be addressed in future investigations. We conclude that the proposed theory offers a coherent phenomenological approach to matter production and provides new insights into non-standard cosmological evolution.


\end{abstract}

\section{Introduction}
\label{S:1}
The study of matter production in cosmological models is an ongoing area of research in modern cosmology and is closely related to fundamental theoretical frameworks. For instance, in steady-state models, the combination of cosmic expansion and the perfect cosmological principle requires the continuous generation of matter. According to Dirac’s hypothesis, the temporal variation of the gravitational constant leads to particle production. Within the thermodynamic theory of irreversible processes, Prigogine et al. aimed to develop a coherent phenomenological approach in which both matter and space-time curvature could be generated simultaneously. This was carried out in the context of the standard hot Big Bang cosmology, with the aim of addressing the matter production problem.

Rastall’s gravity theory is a modification of Einstein’s, in which the covariant divergence of the energy-momentum tensor is not zero but instead proportional to the gradient of the Ricci scalar. This adaptation can effectively describe the consequences of quantum gravity. The non-conservation of the stress-energy tensor, as dictated by the Bianchi identities, leads to new field equations. In this work, we propose a gravitation theory similar to Rastall’s, which conceptually shares the same characteristics but can be derived from a Lagrangian. This framework provides a means to investigate the thermodynamic properties of irreversible processes. In the first section, we present the theoretical background, while section two summarizes the equations of Rastall-like gravity, rewriting them in a form that incorporates the energy-momentum tensor. Section three focuses on the thermodynamics of Rastall-like gravity, while section four analyzes its consequences in the LFRW context for a simple fluid. Finally, section five discusses the results of our approach.

\section{Field Equations}
In standard General Relativity (GR), covariance under general coordinate transformations is a fundamental principle, ensuring that physical laws remain invariant across different reference frames and coordinate systems. This invariance underlies the geometric nature of GR, where the gravitational field, described by the metric tensor  \( g_{\mu\nu} \), is governed by the Einstein field equations, which are covariant under all diffeomorphisms. However, the search for quantum gravity, explanations for dark energy, and modifications on cosmological scales have motivated the exploration of alternative gravitational theories, leading to frameworks that relax or break these symmetries, either locally or globally.

One such alternative is Rastall-like gravity, introduced below, a model that breaks both covariance and the traditional conservation of the energy-momentum tensor, resulting in field equations that deviate from the standard covariant form. In Rastall’s theory, the divergence of the energy-momentum tensor \( T^{\mu \nu} \) is proportional to the gradient of the Ricci scalar, allowing for a modification of energy-momentum conservation in regions where the Ricci scalar varies. This approach challenges conventional conservation laws and redefines their role in gravitational dynamics, particularly in cosmological contexts.

Several gravitational theories either break covariance or modify symmetry principles. Hořava-Lifshitz gravity, for instance, introduces a local violation of Lorentz invariance at high energies to render gravity renormalizable. It distinguishes between temporal and spatial coordinates by introducing a preferred time foliation of spacetime, enforcing invariance under a restricted class of transformations that treat time and space asymmetrically \cite{Horava2009}. This theory has applications in high-energy physics and quantum gravity but fundamentally alters the symmetry properties of spacetime at small scales.

Unimodular Gravity imposes the constraint \( \sqrt{-g} = 1 \), restricting the determinant of the metric tensor and allowing only volume-preserving diffeomorphisms rather than full covariance \cite{Ellis2010}, \cite{Henneaux1989}. This modification enables the cosmological constant to emerge as an integration constant rather than a fundamental parameter, offering a potential resolution to the cosmological constant problem. However, it alters the local degrees of freedom of the gravitational field, affecting the interpretation of conservation laws.

Massive Gravity introduces a mass term for the graviton, necessitating a reference metric that breaks full covariance \cite{deRham2011}, \cite{Hassan2012}. Locally, this reference metric establishes a preferred structure in each region of spacetime, modifying the dynamics by introducing a scale-dependent interaction. This theory has been explored as a possible explanation for cosmic acceleration without invoking dark energy, though a careful formulation is required to avoid instabilities.

Einstein-Æther Theory introduces a dynamic vector field (the “æther”) that breaks Lorentz symmetry by establishing a locally preferred direction in spacetime \cite{Jacobson2008}. Within this framework, the gravitational field equations remain invariant only under transformations that preserve this preferred direction. This approach provides a means to investigate Lorentz symmetry violations and their potential observational consequences, particularly in regions where the æther field strongly influences gravitational dynamics.

In cosmology, theories such as Modified Newtonian Dynamics (MOND) and \( f(R) \)  gravity models modify gravitational interactions on large scales. Certain \( f(R) \) models incorporate functions of the Ricci scalar, altering gravitational dynamics in a scale-dependent manner \cite{Capozziello2008}, \cite{Faraoni2008}. These modifications aim to explain dark matter and dark energy phenomena by introducing variations in the gravitational field across different regions of the universe.

Rastall’s like gravity represents a radical departure from GR’s assumptions, breaking both invariance and the traditional energy-momentum conservation laws. In this framework, the divergence \( \nabla_\mu T^{\mu \nu} \) is proportional to \( \nabla^\nu R \), the gradient of the Ricci scalar. This approach permits variations in energy-momentum exchange with the curvature of spacetime itself, suggesting a fundamental link between matter and geometry that departs from standard conservation principles.

In the context of cosmological modeling, Rastall’s theory introduces an interesting possibility, implying that modified conservation laws allow energy-momentum to evolve dynamically in response to spacetime curvature. This could have significant implications for understanding the large-scale structure of the universe and for testing the evolution of gravitational interactions over time. For instance, Rastall’s approach may offer an alternative explanation for cosmic acceleration by modifying the relationship between energy density, pressure, and curvature in expanding regions of spacetime.

While these modifications to invariance and conservation provide flexibility in gravitational theory, they also pose significant theoretical and observational challenges. Full covariance and conservation serve not only as symmetry principles but also as foundational tools in defining physical laws across regions of spacetime. Theories that relax or break these principles, such as Rastall’s like theory, must ensure internal consistency and empirical viability across different scales or regions.

Furthermore, the local and global breaking of covariance can complicate the interpretation of conserved quantities and lead to ambiguities in defining energy or momentum. In Rastall’s like theory, the non-conservation of energy-momentum could impact the standard formulation of cosmological equations, necessitating new methods to ensure compatibility with observational data, particularly in regions where curvature gradients are significant.

The exploration of gravitational theories that break full invariance and covariance, such as Rastall’s like theories, provides intriguing possibilities for rethinking gravitational interactions. Although such frameworks challenge traditional conservation laws and symmetry principles, they open pathways for addressing longstanding cosmological puzzles, such as dark energy and cosmic acceleration. This integration of Rastall’s theory into the discussion particularly highlights how its breaking of both invariance and conservation could provide a novel approach in cosmological modeling and testing.

\section{Lagrangian type Rastall} \label{sec:1}
We begin by proposing an initial Lagrangian for Rastall-like gravity, which is expressed as follows:
\begin{eqnarray}
{\cal S} = \int d^4 x \sqrt{-g} \Big[ R- \xi(g) \; R+ \kappa \; {\cal L}_m \Big].
\end{eqnarray}

The field equations for Rastall-like gravity are given by:
\begin{eqnarray}\label{FieldEquation}
\left(1-\xi(g) \right) R_{\mu\nu}-\frac{1}{2} g_{\mu\nu} R= \kappa T_{\mu\nu},
\end{eqnarray}
where $\xi(g)=\beta/\sqrt{-g}$, with $\beta$ being a constant and ${\cal L}_m$ the matter Lagrangian. 

When $\beta=0$, the Einstein equations are recovered. The Bianchi identities remain satisfied; however, the conservation of the stress-energy tensor for the gravity source is violated, as given by:
\begin{eqnarray}\label{NonEnergyConservation}
\kappa\nabla_{\mu} T_{\nu}^{\mu}=-\frac{\xi(g) }{2} \; \nabla_{\nu} R.
\end{eqnarray}

Here, $\kappa$ is a dimensional constant that must be determined to ensure the correct Poisson equation in the static weak-field limit.

We now focus on applying this framework in cosmology. First, we investigate the thermodynamic effects in the context of an ideal gas. Then, we proceed to analyze the LFRW model.
 
\section{Thermodynamics and matter production for a simple fluid in Rastall like gravity}
We can think of the universe as an open thermodynamic system in which particles are produced due to the gravitational field, as we will demonstrate in this section. This is allowed by Rastall-like gravity. From now on, we assume spatial isotropy, and in this work, we will restrict our considerations to an energy-momentum tensor of the form $ T_{ab} =\rho v_{a} v_{b} -p h_{ab}$, 
which, in Rastall-like gravity, satisfies the conservation law given by Eq. (\ref{NonEnergyConservation}).  Here $\rho$ is the energy density, $p$ is the equilibrium isotropic pressure, and $v^{a}$  is the fluid four-velocity. This vector is time-like, given by $v_{a}v^{a}=1$. The projection tensor onto the subspace orthogonal to $v^{a}$, is defined as $$
h_{ab}= g_{ab}-v_{a}v_{b},$$
and satisfies the following relations:  $$v^{a}h_{ab}=0,\;  h_{ab}h^{ab}=3,\; \mbox{and}\; h_{\;a}^{c}h_{\;b}^{a}=\delta^{c}_{b}.$$ \\

The macroscopic variables describing the thermodynamic states of a relativistic simple fluid are the energy-momentum tensor $T_{ab} $, the particle flux vector $N^{a}$, and the entropy flux vector $S^{a}$. The energy-momentum tensor $T_{ab} $  was described earlier. The particle flux vector is assumed to take the form: $$N^{a}=n\;v^{a}$$ where $n$ is the particle number density. This vector satisfies the balance equation
\begin{equation}\label{PFluxB}
 \nabla_{a}N^{a}= {\cal J}_n,
\end{equation}
\noindent where ${\cal J}_n$ is a particle source if ${\cal J}_n>0$ or as a sink if ${\cal J}_n<0$. 

The entropy flux vector is given by $$S^{a}=n\sigma v^{a},$$ where $\sigma$ is the specific entropy per particle. This quantity satisfies the second law of thermodynamics:
\begin{equation}\label{SL}
 \nabla_{a} S^{a}\geq 0 .
\end{equation}

The Euler equation is given by $$\mu =\frac{\rho+p}{n} -T \sigma,$$
where $T$ is the temperature and $\mu$ is the chemical potential. For this system, the Gibbs-Duhem equation reads:
$$ n T d\sigma=d\rho-\frac{\rho+p}{n} \; dn.$$
The equations (3), (\ref{PFluxB}), and (\ref{SL}), together with Euler and Gibbs-Duhem equations provide the full set of thermodynamic laws in general relativity with particle production.

Now, we derive the thermodynamic equations in the context of Rastall-like gravity. To this end, we assume a continuous energy transfer from the variation of curvature to the material component described by  $\rho$.  Consequently, the energy conservation, eq.(\ref{NonEnergyConservation}), projected  along $v_{a}$  for a perfect fluid takes the form:
\begin{equation} \label{nd}
\dot{\rho} +  ( \rho+p ) \theta  =  - \lambda(g)  \dot{R},
\end{equation}
where $\lambda(g)=\lambda=\frac{\beta }{2\kappa\sqrt{-g}}$ the overdot denotes the covariant  derivative along the worldlines, such that $\dot{\rho}\equiv v^{a} \nabla_{a} \rho $ , $\theta\equiv \nabla_{a}v^{a}$ is the expansion scalar.

To complete the fluid description, we use equation (\ref{PFluxB}):
\begin{equation}\label{np}
\dot{n}+n\theta = {\cal J}_n,
\end{equation}
which describes the evolution and production of the particle number density. Finally from equation (\ref{SL}), the second law of thermodynamics, combined with the Gibbs-Duhem and Euler equations, leads to
\begin{eqnarray} \label{fe}
\nabla_{a} S^{a}= -\lambda \frac{\dot{R}}{T}-\frac{\mu}{T}{\cal J}_n \geq 0 .
\end{eqnarray}

From this result, we identify the covariant Gibbs-Duhem equation : 
\begin{eqnarray}
T \nabla_{a} S^{a}= v_{\alpha} \nabla_{\beta} T^{\alpha\beta} -\mu \nabla_{\alpha}N^{\alpha} .
\end{eqnarray}
To carry out the thermodynamic analysis, we must consider the temperature evolution. We begin by assuming $\rho=\rho (T,n)$, since $\rho$ is an exact differential. By utilizing (\ref{nd}) and (\ref{np}), it is straightforward to determine that

\begin{equation}\label{rho}
\left. \frac{\partial \rho}{\partial T} \right)_{n} \dot{T}= \left[ n\left. \frac{\partial \rho}{\partial n} \right)_{T}-(\rho+p)\right] \theta \; -   \left. \frac{\partial \rho}{\partial n} \right)_{T} {\cal J}_n -\lambda \dot{R} ,
\end{equation}
where the first term on the right-hand side represents the usual equilibrium contribution of this law. The second term is associated with particle production, and the non-conservation of matter is a feature of Rastall's gravity theory. 

\noindent By combining $d\sigma(n,T)$ with the Gibbs-Duhem equation, we can derive a temperature evolution law: 
\begin{equation}
\frac{\dot{T}}{T}= \frac{\dot{n}}{n}   \left. \frac{\partial p}{\partial \rho} \right)_{n} - \frac{1}{n \;T  \left. \frac{\partial \rho}{\partial T} \right)_{n}} \left[(\rho+p){\cal J}_n+\lambda n \dot{R}  \right],
\end{equation}
where $C_s=\left. \frac{\partial p}{\partial \rho} \right)_{n}$ is the adiabatic speed of sound. 

\noindent In the next section we particularise the metric. 

\section{LFRW-type cosmologies}
The Lemaître-Friedmann-Robertson-Walker (LFRW) metric describes spacetime under the assumptions of homogeneity and isotropy of space: 
\begin{equation}
ds^2 = dt^2 - a(t)^2\biggr[dx^2 + dy^2 + dz^2\biggl]\;.
\end{equation}
 where we consider a spatially flat LFRW universe with $k = 0$. The energy-momentum tensor takes the form $ T_{ab} =\rho v_{a} v_{b} -p h_{ab}$. Then, equation (2)  yields:

\begin{eqnarray} 
\kappa\rho=&& \frac{3}{a(t)^2}\left(\frac{\partial a(t)}{\partial t}\right)^2 + \frac{3\beta}{a(t)^4} \frac{\partial^2 a(t)}{\partial t^2} \label{FriedmanE1}\;, \\
\kappa p=&& -\frac{1}{a(t)^2}\left(\frac{\partial a(t)}{\partial t}\right)^2-  \frac{2\beta}{a(t)^5} \left(\frac{\partial a(t)}{\partial t}\right)^2\\&&-\frac{\beta}{a(t)^4}\frac{\partial ^2 a(t)}{\partial t^2}\nonumber  -\frac{2}{a(t)}\frac{\partial^2 a(t)}{\partial t^2}\label{FriedmanE2}\;, 
\end{eqnarray} 
where the function $a(t)$  is the scale factor, describing the expansion of the universe, $\rho$  is the energy density,  and $p$  is the corresponding pressure.   The Ricci scalar curvature of space is given by:

\begin{equation*}
 R={\frac {6}{a^{2}}}\left[a \frac{\partial^2 }{\partial t^2}a+\left(\frac{\partial a}{\partial t}\right)^2\right],
\end{equation*} which, in terms of the Hubble parameter $H(t)=\dot a/a$, becomes: 
\begin{equation}
    R=6 \left(\dot H  +2 H^2\right).
\end{equation}
The deceleration parameter $q$ is a dimensionless quantity  in cosmology  that measures the acceleration of the universe’s expansion, defined as $q=-a\ddot a /\dot a ^2 $. From Eq.(13) and (14), we obtain:
\begin{equation}
    q=\frac{(w+\frac{1}{3})a^3-\frac{2}{3}\beta}{\frac{2}{3} a^3+\beta(w+\frac{1}{3})}=-\frac{a\ddot a}{ \dot a ^2}.
\end{equation}
Where we  introduce the parameter $w$, defined as the equation-of-state parameter:
\begin{equation}
    w=\frac{p}{\rho}=-\frac{1}{3}-\frac{2}{3} \left( \frac{a^4\ddot a+\beta \dot a^2}{\beta a\ddot a+a^3\dot a^2}\right).
\end{equation}
The dynamical evolution of physical quantities, such as the equation-of-state parameter $w$, depends on the evolution of the scale factor $a(t)$ and the parameter $\beta$.  In the limit $\beta=0$, the model reduces to General Relativity.\\  
\noindent In the next section, we analyze the role of particle production in enabling a bounce scenario.\\

\section{Cosmological Bouncing Solutions in Rastall-like Gravity theories }
The concept of a cosmological bounce presents an alternative to the standard Big Bang model. It postulates that the universe undergoes a significant contraction followed by a rebound, rather than originating from an initial singularity. In this framework, the universe evolves through cyclic or oscillatory phases, where each cosmological cycle begins with the collapse of a previous phase. Such a scenario not only offers a potential resolution to the singularity problem inherent in standard cosmology, but also serves as an alternative to the inflationary paradigm. This work focuses on analyzing bouncing cosmological models, with particular emphasis on the role of particle production within the context of Rastall-like gravity theories.\\
A successful bouncing scenario in standard cosmology must satisfy certain conditions. During the contraction phase, the scale factor satisfies $a(t)<0$ , while during the expansion phase, $a(t)>0$. At the bounce point, the scale factor reaches a non-zero minimum value, and $\dot{a}(t)=0$, with $\ddot{a}(t)>0$  in its vicinity. This behavior naturally avoids the singularity that is unavoidable in conventional Big Bang cosmology.\\
Furthermore, the Hubble parameter $H(t)$ transitions from negative values ($H(t)<0$) in the contracting phase to positive values ($H(t)>0$) in the expanding phase, with $H(t)=0$ exactly at the bounce point. In standard general relativity, for a bounce to occur, it is required that $\dot{H}(t)>0$ near the bounce. This condition implies a violation of the null energy condition (\textbf{NEC}), typically associated with an equation-of-state parameter satisfying $w < -1$ in the neighborhood of the bounce.\\
In this study, we consider a specific functional form of the scale factor $a(t)$, and investigate the corresponding behavior of the equation-of-state parameter $w$. Our analysis is particularly focused on scenarios where $w>-1$ in the recent past and transitions to $w<-1$ in the present, aiming to understand the conditions under which a bounce can be achieved in the framework of Rastall-like gravity.

\subsection{Dynamical analysis of the bouncing evolution}
We now focus on the behavior of the Hubble parameter $H(t)$, or equivalently, the scale factor $a(t)$, which characterizes the expansion dynamics of the universe. Our goal is to demonstrate how these quantities can lead to significant bouncing solutions within the framework of the modified Friedmann equations in Rastall-like gravity. In particular, we consider a specific parametrization for the scale factor during the bouncing phase, given by:
\begin{equation}
a(t)=\alpha \left(1+\frac{ t^2}{l^2}\right)^{\frac{1}{n}},
\end{equation}
where $\alpha>0$,$l>0$, and $n$ are constants. This form ensures that $a(t)$ remains finite and strictly positive, with a minimum at $t=0$, thus satisfying the basic condition for a bounce.\\
The corresponding Hubble parameter is: 
\begin{equation}
    H(t)=\frac{2t}{n(l^2+t^2)},
\end{equation}
which clearly vanishes at the bounce point $t=0$, changing sign across it — a hallmark of a bouncing universe.
The deceleration parameter, which indicates whether the expansion is accelerating or decelerating, is given by:
\begin{equation}
    q(t)=\frac{1}{2}\left[n(1-\frac{l^2}{t^2})-2\right] ,
\end{equation}
 revealing a rich behavior near the bounce, particularly around $t\approx 0$, where acceleration becomes dominant.
Finally, the Ricci curvature scalar associated with the LFRW geometry is expressed as:
\begin{equation}
    R=\frac{12 \left[l^2 n - (n - 4) t^2\right]}{n^2 (l^2 + t^2)^2}, 
\end{equation}
which describes a non-singular curvature evolution of spacetime through the bounce phase.\\
From this, we observe that the Hubble parameter and the Ricci scalar curvature vanish both at early and late times, while the deceleration parameter remains finite throughout the evolution. At the bouncing point, $H(t) = 0$, the deceleration parameter $q(t)$ diverges, whereas the Ricci scalar curvature remains finite, approaching $R \rightarrow  \frac{12}{n l^2}$. It is interesting to note that \( l^2 \) is  inversely related to the curvature: as \( l^2 \) becomes very small—on the order of the Planck length squared—the curvature increases significantly and tends toward infinity as \( l^2\rightarrow 0 \), similar to what occurs in the standard Friedmann model. Furthermore, it is worth emphasizing that this parameter \( l^2 \) appears explicitly in the expression for the scale factor.

\begin{figure}[htbp] 
    \centering
    \begin{subfigure}{0.45\textwidth}
        \centering
        \includegraphics[width=\linewidth]{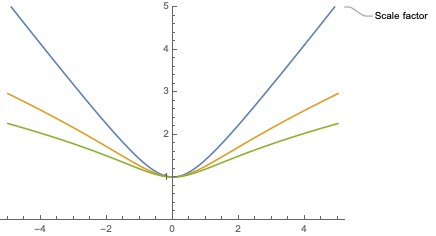}
      
        \label{fig:fig1}
    \end{subfigure} 
    \hfill
    \begin{subfigure}{0.45\textwidth}
        \centering
        \includegraphics[width=\linewidth]{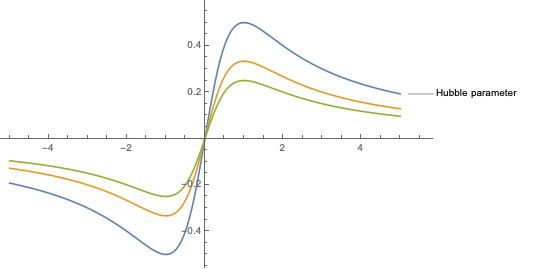}
        
        \label{fig:fig2}
    \end{subfigure}

    \vspace{0.5cm}
    \begin{subfigure}{0.45\textwidth}
        \centering
        \includegraphics[width=\linewidth]{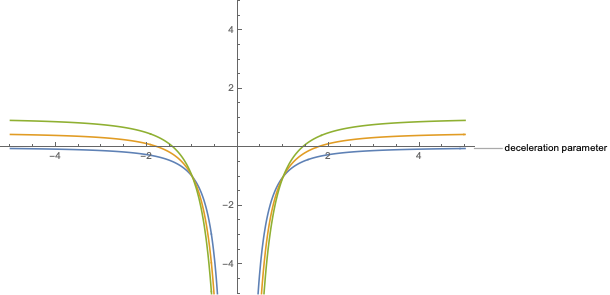}
        
        \label{fig:fig3}
    \end{subfigure}
    \hfill
    \begin{subfigure}{0.45\textwidth}
        \centering
        \includegraphics[width=\linewidth]{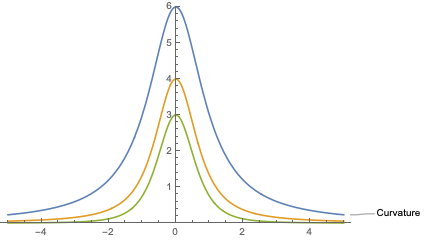}
        
        \label{fig:fig4}
    \end{subfigure}

    \caption{Behavior of the scale factor, Hubble parameter, deceleration parameter, and Ricci scalar curvature as functions of cosmic time. The blue curve corresponds to \( n = 2 \), the orange curve to \( n = 3 \), and the green curve to \( n = 4 \).}
    \label{fig:fig_grid1}
\end{figure}
Since $a$, $l$, and $n$ are positive constants, we fix their values as $a=1$, $l=1$, and $n = 2, 3, 4$ in order to analyze the behavior of the cosmological parameters. In Fig. \ref{fig:fig_grid1}, the plots of the scale factor, Hubble parameter, deceleration parameter, and Ricci scalar curvature exhibit behavior consistent with the conditions required for a successful bounce. The good agreement between the analytical predictions and the numerical behavior of these parameters reinforces the validity and predictive capability of the model.

Specifically, at the bouncing point, the scale factor attains its minimum value before beginning to increase, as illustrated in Fig. \ref{fig:fig_grid1}. The Hubble parameter initially assumes negative values, corresponding to a contracting phase ($H < 0$). It then vanishes at the bounce point ($H = 0$) and subsequently becomes positive, signaling the onset of an expanding phase ($H > 0$). 

The behavior of the deceleration parameter diverges at the singularity corresponding to the bouncing point and approaches the asymptotic value  $q = (n - 2)/2$ at late times. Furthermore, the Ricci curvature scalar reaches its maximum value at the bouncing point. These dynamics highlight the significant transitions in the Universe's evolution from contraction to expansion, capturing essential features of the bouncing cosmological model under consideration.
In this context, the equation-of-state parameter $w$ is given by:
\begin{equation}
 w=\frac{1}{3}-\frac{2}{3}  \frac{2 \beta t^2+\left(n-n t^2+2 t^2\right) \left(t^2+1\right)^{3/n}}{\beta \left(n-n t^2+2 t^2\right)+2 t^2 \left(t^2+1\right)^{3/n}}.
\end{equation}
As mentioned previously, the $w$ parameter plays  a fundamental role in understanding the dynamics of the Universe, especially during key cosmological events such as the bouncing phase. In bouncing models, $w$ governs the relation between the energy density and pressure of the cosmic fluid, thereby influencing the conditions required for a non-singular transition between the contracting and expanding phases.

When $\beta \rightarrow 0$, which corresponds to the case of General Relativity, the equation-of-state parameter $w$ behaves as:  
\begin{equation}
w = \frac{-l^2 n}{3 t^2} + \frac{(n - 3)}{3},
\end{equation}
At late times, this expression asymptotically approaches: 
\begin{equation}
w = \frac{1}{3}(n - 3).
\end{equation}

We observe that this result is consistent with the standard cosmological cases of dust ($w=0$) and radiation ($w=1/3$), which correspond to $n = 3$ and $n = 4$, respectively. This behavior imposes a restriction on the parameter $n$, requiring that $ n \geq 3$ to yield physically meaningful values at late times.

At $t = 0$ and and in the limit $\beta \rightarrow 0$, the parameter $w$ diverges, as illustrated in Fig. 2. However, when $\beta \neq 0$ and $t = 0$, we obtain:
\begin{equation}
w = \frac{1}{3}-\frac{2 }{3 \beta} .
\end{equation}

This result is particularly significant, as it allows for the presence of exotic components such as a quantum fluid. In this scenario, the \( w \) parameter can take negative and finite values—for instance, it may reach as low as \(w= -1 \). Notably, all physical quantities within the bouncing model remain well-behaved and consistent, even at the bounce.

\begin{figure}[ht] 
    \centering
    \begin{subfigure}{0.45\textwidth}
        \centering
        \includegraphics[width=\linewidth]{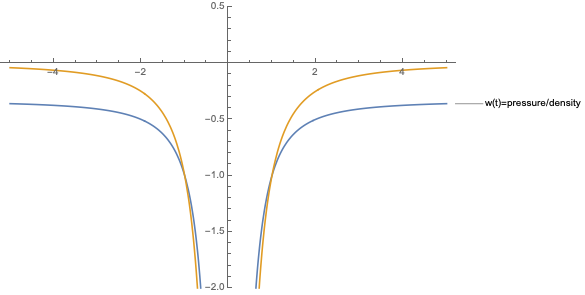}
      
        \label{fig:fig5}
    \end{subfigure} 
    \hfill
    \begin{subfigure}{0.45\textwidth}
        \centering
        \includegraphics[width=\linewidth]{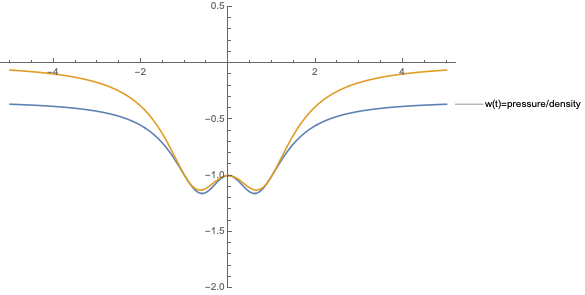}
        
        \label{fig:fig6}
    \end{subfigure}

    \caption{The behavior of the equation-of-state parameter $w$ as a function of cosmic time for the cases $\beta=0$ and $\beta\neq 0$.  In the figure, the blue curve corresponds to $n = 2$ and the orange curve to $n = 3$}
    \label{fig:fig_grid2}
\end{figure}
\subsection{Violation of Energy Conditions and Stability Analysis }
Energy conditions provide fundamental constraints on various linear combinations of energy density and pressure. They impose requirements such as the non-negativity of energy density and the attractive nature of gravity. These conditions emerge from the Raychaudhuri equation and play a pivotal role in assessing the physical viability of gravitational models. The main energy conditions—the Null Energy Condition (NEC), Weak Energy Condition (WEC), Dominant Energy Condition (DEC), and Strong Energy Condition (SEC)—are defined as follows:

\begin{itemize}
    \item[$\blacksquare$] \textbf{NEC:} $  \rho + p \geq\, 0 $
\item[$\blacksquare$] \textbf{WEC:} $\rho \geq 0 \quad \text{and} \quad \rho + p \geq 0 $  
\item[$\blacksquare$] \textbf{DEC:} $ \rho \geq 0 \quad \text{and} \quad |p| \leq \rho$  
\item[$\blacksquare$] \textbf{SEC:} $ \rho + 3p \geq\, 0$
\end{itemize}
\noindent To establish a viable bouncing cosmological scenario, it is essential to examine the energy conditions in detail to ensure that the model remains physically consistent and effectively captures the transition between the contracting and expanding phases of the Universe.

\noindent As illustrated in Fig. 3a, a violation of the \textbf{NEC} \((\rho + p < 0)\) occurs at the bouncing point. This violation is not only expected but also necessary, as it enables the bounce by momentarily altering the standard gravitational behavior. Thus, the \textbf{NEC} violation confirms that the model satisfies the fundamental requirements for realizing a bounce, further supporting the theoretical soundness of the proposed cosmological framework.

\noindent Moreover, our analysis shows that the  \textbf{SEC} \((\rho + 3p > 0)\) is violated near the bouncing point at \(t = 0\), suggesting the presence of exotic matter or non-standard energy components. These components are essential to counteract gravitational collapse and enable the bounce. In contrast, the  \textbf{DEC} \((\rho - p \geq 0)\) remains satisfied, ensuring that the energy density continues to dominate over pressure, which is crucial for preserving physical consistency.

\noindent Hence, the proposed model fulfills all the essential criteria for a consistent bouncing cosmology, particularly within the framework of a Rastall-type gravitational theory. These results reinforce the model’s viability and underscore its potential as a compelling alternative to the conventional Big Bang scenario.

\begin{figure}[htbp] 
    \centering
    \begin{subfigure}{0.45\textwidth}
        \centering
        \includegraphics[width=\linewidth]{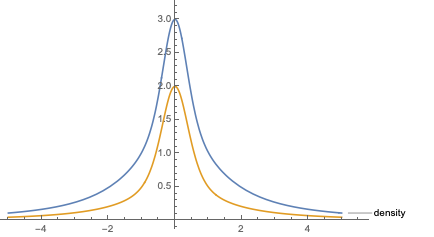}
      
        \label{fig:fig7}
    \end{subfigure} 
    \hfill
    \begin{subfigure}{0.45\textwidth}
        \centering
        \includegraphics[width=\linewidth]{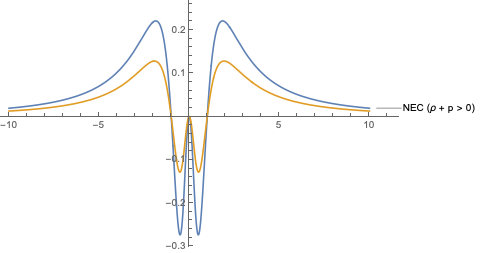}
        
        \label{fig:fig8}
    \end{subfigure}

    \vspace{0.5cm}
    \begin{subfigure}{0.45\textwidth}
        \centering
        \includegraphics[width=\linewidth]{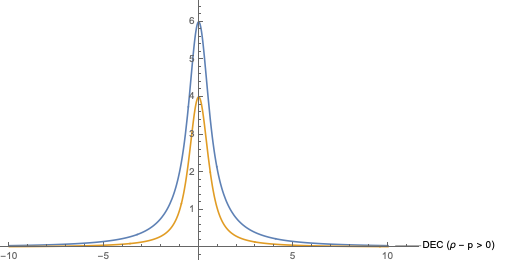}
        
        \label{fig:fig9}
    \end{subfigure}
    \hfill
    \begin{subfigure}{0.45\textwidth}
        \centering
        \includegraphics[width=\linewidth]{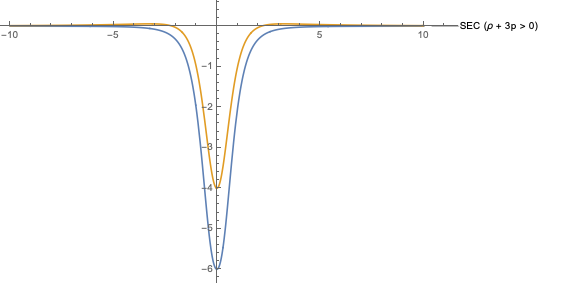}
        
        \label{fig:fig10}
    \end{subfigure}
\caption{Evolution of the energy conditions as functions of cosmic time for the case $w = -1$. The blue curve represents $n = 2$, while the orange curve corresponds to $n = 3$. The model exhibits a violation of the NEC and SEC at the bouncing point, a fundamental requirement to avoid gravitational collapse and enable the transition. The validity of the DEC ensures the energy density dominates over pressure, preserving the physical consistency of the cosmological framework.}    
    \label{fig:fig_grid3}
\end{figure}
In the following analysis, we investigate the stability of the model by considering a universe filled with a perfect fluid, where the adiabatic speed of sound, denoted as \( C_s \), plays a key role. As defined after Equation (11), this parameter is crucial for understanding the behavior of cosmological fluids. The speed of sound characterizes the velocity at which perturbations (sound waves) propagate through the medium, which in cosmological scenarios is typically composed of baryonic matter such as gas and plasma. The speed of sound is mathematically defined by:
\[
C_s=\left. \frac{\partial p}{\partial \rho} \right)_{n} , 
\]
where \(p\)  is the pressure, \(\rho\) is the energy density of the fluid, and the derivative is taken at constant entropy (adiabatic condition). This expression reflects how changes in pressure relate to changes in density, offering insight into the responsiveness of the fluid to perturbations and, therefore, into the overall stability of the cosmological model.

The constraints on \(C_s^2\) are governed by two fundamental limits. The lower bound, \(C_s^2 \geq 0\), ensures the stability of the medium—a negative value would imply that pressure decreases as density increases, leading to unphysical instabilities in the fluid. Therefore, a non-negative speed of sound is a necessary condition for a stable cosmological model. The upper bound, \(C_s^2 \leq 1\), stems from the relativistic principle that no signal or perturbation can propagate faster than the speed of light in a vacuum  (\(c\)).  In cosmological contexts, it is standard to normalize the speed of sound using natural units where $c=1$, yielding the condition \(0\leq C_s^2 \leq 1\).

\noindent This bounded behavior of \(C_s^2\) plays a crucial role in analyzing the evolution of density fluctuations, the formation of large-scale structures, and the propagation of acoustic waves in the early universe. In particular, it directly impacts our understanding of processes such as baryon acoustic oscillations and the emergence of galaxies and cosmic web structures.
\begin{figure}[h] 
    \centering 
    \includegraphics[width=0.8\textwidth]{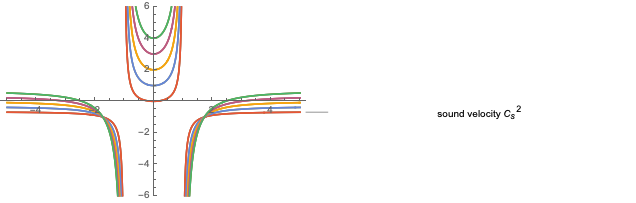} 
   \caption{Cosmic time evolution of the sound velocity for increasing values of $n$ ($n = 1$ to $5$), represented by curves ranging from red to green.}

    \label{fig:minha_figura4} 
\end{figure}

In our  model, the squared sound speed \( C_s^2 \) is given by: 
\begin{equation}
   C_s^2= \frac{2 a^3 \left((n-3) t^2-l^2 n\right) \left(\frac{t^2}{l^2}+1\right)^{3/n}+b \left((n-6) t^2-l^2 n\right)}{6 a^3 t^2 \left(\frac{t^2}{l^2}+1\right)^{3/n}+3 b l^2 n-3 b (n-2) t^2}.
\end{equation}
In Fig. 4, we present the stability analysis of the modified Rastall cosmological model. Near the bouncing point, the stability condition is satisfied for \( n = 1 \) and \( n = 2 \); however, for other values of $n$, this condition is not fulfilled. Moreover, in the late-time evolution, the stability condition remains violated—as $ C_s^2$ stays negative—except for the cases \( n = 4 \) and \( n = 5 \). These findings suggest that the model exhibits some stability at early times for lower $n$, while for \( n = 4 \) and \( n = 5 \), stability is achieved in the later stages of cosmic evolution. This observation highlights the need for a deeper investigation into the specific dynamics of the model.

\begin{figure}[h] 
    \centering 
    \includegraphics[width=0.8\textwidth]{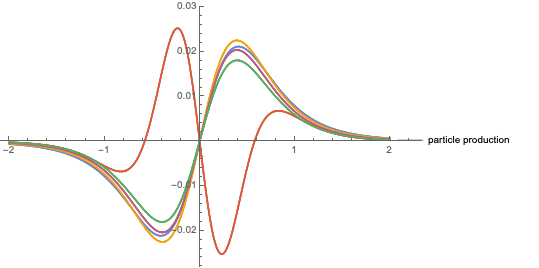} 
    \caption{Evolution of the sound velocity as a function of cosmic time for $n = 1$ to $5$, represented by curves ranging from red to green.}

    \label{fig:minha_figura5} 
\end{figure}

We now examine the particle production generated within the framework of the Rastall model, which emerges due to the curvature scalar $R$. Starting from Eq. (8), we find that entropy production vanishes, leading us to identify the particle production source term as  \({\cal J}_n = -\lambda \dot R / \mu\). Using the definition \(\lambda = \frac{\beta}{2\kappa\sqrt{-g}}\), we obtain \({\cal J}_n = -\text{constant} \cdot \dot R / a(t)^3\), where the scale factor $a(t)$ is explicitly introduced, and the constant can either be absorbed into $\beta$ or normalized in the source term.

\noindent This yields the following expression:
\begin{equation}
   {\cal J}_n= -\frac{24 b t \left(\frac{t^2}{l^2}+1\right)^{-3/n} \left(l^2 (4-3 n)+(n-4) t^2\right)}{a^3 n^2 \left(l^2+t^2\right)^3},
\end{equation}
figure 5 displays the numerical solutions for different values of $n$, showing that particle production consistently peaks at the bouncing point regardless of the choice of $n$. This provides a natural mechanism for particle production within the theory, without affecting the late-time evolution. Moreover, as indicated by Eq. (11), the temperature remains unaffected as long as the stability condition—previously discussed—is satisfied.

\section{Conclusion}

In this work, we have explored the thermodynamic implications and cosmological applications of Rastall-like gravity, offering a modified theoretical framework in which particle production emerges naturally from deviations in the conventional conservation of the energy-momentum tensor. Below, we summarize the main findings and their relevance to cosmological theory.

In Rastall-like gravity, the energy-momentum tensor is not strictly conserved, allowing for continuous matter production driven by the gravitational field. This feature is consistent with the thermodynamics of open systems, where non-conservation leads to particle creation. Our analysis shows that, under specific conditions, the entropy per particle remains constant—indicating that the model can approximate adiabatic processes and maintain thermodynamic equilibrium in certain regimes. This property makes the theory particularly suitable for investigating the early universe, where particle production plays a crucial role in cosmic evolution.

A significant outcome of our study is the development of a bouncing cosmological model. Unlike the standard Big Bang scenario, in which the universe originates from a singularity, the bouncing model envisions a cyclic process, characterized by successive phases of contraction and expansion. The existence of a non-zero minimum for the scale factor at the bounce point effectively resolves the singularity problem that plagues conventional cosmology. In our model, the Hubble parameter evolves smoothly from negative values—associated with the contracting phase—to positive values—corresponding to expansion—thereby confirming the dynamical viability of the bounce. The violation of the null energy condition (NEC) near the bounce is a necessary and well-motivated feature of this scenario, as it temporarily reverses standard gravitational behavior and permits the transition to expansion. Notably, the Rastall-like framework naturally accommodates such a violation without undermining the internal consistency of the model, further reinforcing its potential as a robust alternative to singular cosmologies.

Our stability analysis, based on the adiabatic speed of sound, indicates that the model remains stable under specific conditions during both the early and late phases of cosmic evolution. The energy conditions—null, weak, dominant, and strong—serve as essential criteria for assessing the physical consistency of any gravitational theory. While the null and strong energy conditions are violated near the bounce point, these violations are not pathological but rather necessary to enable the transition from contraction to expansion. Importantly, the weak and dominant energy conditions remain satisfied throughout the evolution, preserving the model’s physical viability.

The curvature-induced particle production mechanism, derived from the thermodynamic equations of the Rastall-like framework, offers a compelling and self-consistent explanation for matter generation. Unlike conventional models that require external sources to account for particle creation, Rastall-like gravity inherently couples this process to the dynamics of spacetime itself. This feature presents a natural alternative to inflationary scenarios and opens new avenues for studying the early universe beyond the standard paradigm.

Moreover, our results show that the entropy produced during particle creation has negligible effects on the temperature evolution at late times, ensuring compatibility with current cosmological observations.

This study lays the groundwork for future investigations into non-equilibrium thermodynamic processes in cosmology. Further research may involve more detailed numerical simulations to confront the model’s predictions with data from cosmic microwave background (CMB) anisotropies and large-scale structure formation. It is also worthwhile to explore how variations in the scale factor or particle production rates affect the universe’s dynamics. Additionally, establishing connections between Rastall-like gravity and quantum gravity approaches could yield deeper insights into the early universe and the fundamental nature of gravity.

Finally, this theoretical framework offers a fertile ground for studying exotic matter fields and their role in stabilizing bouncing cosmologies. Such inquiries may enhance our understanding of the intricate relationship between matter production, gravitational dynamics, and thermodynamic behavior in non-standard cosmological models.

In summary, Rastall-like gravity provides a coherent and flexible platform for addressing some of the most profound challenges in cosmology, including the singularity problem and the origin of matter. The thermodynamic processes embedded in this theory allow for novel insights into the production of particles and the overall evolution of the universe. The bouncing cosmological scenario presented here emerges as a promising alternative to the inflationary paradigm, offering physically consistent and mathematically robust solutions. Although this work is purely theoretical, the results establish a solid foundation for future studies aiming to confront the model with cosmological data, including large-scale structure formation and CMB observations. As interest in alternative gravitational theories continues to grow, the results presented in this work underscore the potential of Rastall-like gravity to advance our understanding of the universe’s origins and its large-scale behavior.



\end{document}